\documentclass[preprint,12pt]{elsarticle}

\usepackage{amssymb}

\journal{---}

\begin{document}

\begin{frontmatter}



\title{On de Broglie's quantum particle as the soliton solution of linear Schr\"odinger equation}


\author{Agung Budiyono}

\address{Institute for the Physical and Chemical Research, RIKEN, \\2-1 Hirosawa, Wako-shi, Saitama 351-0198, Japan}

\begin{abstract}

We develop a class of soliton solution of {\it linear} Schr\"odinger equation without external potential. The quantum probability density generates its own boundary inside which there is internal vibration whose wave number is determined by the velocity of the particle as firstly conjectured by de Broglie. Assuming resonance of the internal vibration will lead to quantization of particle's momentum in term of wave number of the envelope quantum probability density. We further show that the linearity of the Schr\"odinger equation allows us to have non-interacting many solitons solution through superposition, each describing a particle with equal mass. 

\end{abstract}


\end{frontmatter}


\section{Introduction: de Broglie's double solution}

In his attempt to explain the dual nature of matter as both wave and particle, de Broglie proposed what is called as the theory of double solution \cite{de Broglie book}. He was searching for a nonlinear type of wave equation  which possesses a singular, spatially localized, nondispersive  solution in region where the amplitude is sufficiently high. This spatial part of the global solution is then supposed to play the role of a particle. On the other hand,  in region where the amplitude is weak, the solution is assumed to behave like a linear wave obeying the law of superposition. He thus envisioned a particle which is immersed in linear wave field and is guided by the latter. He argued that in order that each point of the singular solution to move non-dispersively, thus particle-like, the velocity of the singular solution must be equal to the velocity of the non-singular linear solution. This is de Broglie's guiding principle \cite{de Broglie book} which eventually led him to the famous relation between the momentum of a particle $p$ and the associated wave  length $\lambda$ of the guiding wave:
\begin{equation}
p=\hbar k,
\label{de Broglie relation}
\end{equation}
where $k=2\pi/\lambda$ is the corresponding wave number. 

The prevailing quantum theory is however founded on the Schr\"odinger equation which is linear with respect to its ingredient, namely the wave function. In this theory, a single free particle is usually represented by a plane wave. Yet, the successes of this representation seems to rely on the adoption of Eq. (\ref{de Broglie relation}) where $k$ is now the wave number of the plane wave,  and the relation between the energy of the particle and the frequency of the plane wave,  $\omega$: $E=\hbar\omega$ \cite{de Broglie late book}. One can even argue that those two relations are the most important principles of quantum theory \cite{Pauli book,Barut paper}. Even the linear Schr\"odinger equation can be derived from those relations. 

It is then apparently the linearity of the Schr\"odinger equation which discourages people to keep searching for particle-like solution.  The ambition to search for a particle-like solution is further overshadowed by the tremendous successes  (for all pragmatical purposes \cite{Bell unspeakable}) of the probability interpretation of wave function invented by Born \cite{Born paper} which is heavily based on the law of superposition.  This law is possible only if the underlying equation is linear. This eventually led to the axiomatic development of quantum theory by Dirac and von Neumann \cite{Dirac book,von Neumann book}. 

Despite of these facts, fueled by the dissatisfaction on the foundation of the axiomatic approach to quantum theory \cite{Bell unspeakable,Isham book}, and a simple fact from, say the single slit experiment which undeniably suggests that there is indeed a localized quantity which hits the screen thus giving a point of scintillation \cite{Bell unspeakable,Bohm-Hiley book}, some people are still searching for  a quantum theory which assumes a localized particle-like solution. Interestingly, this adventure is again led by de Broglie after abandoning his own idea for many decades. Much effort  again is made to search for such a solution from a nonlinear wave equation \cite{Curfaro-Petroni paper,Vigier paper 1,Vigier paper 2, Mackinnon paper 1}. 

In this paper, we shall show that even a linear Schr\"odinger equation possesses a class of localized-nondispersive solution which interestingly shares the properties long conjectured by de Broglie. This will be done by entertaining the hydrodynamics picture of the Schr\"odinger equation developed by Madelung \cite{Madelung paper}. Even more, we show that the soliton wave function is the limiting case of a class of the most probable wave functions given its quantum energy \cite{AgungPRA1}. We then show that the linearity of the Schr\"odinger equation allows one to develop many solitons solution through superposition. 

\section {Most probable quantum probability density} 

Let us consider the following {\it linear} Schr\"odinger equation which is supposed to describe a single free particle with mass $m$
\begin{equation}
i\hbar\partial_t\psi(q;t)=-\frac{\hbar^2}{2m}\partial_q^2\psi(q;t).
\label{Schroedinger equation}
\end{equation}
Here $q$ is space and $t$ denotes time. For our discussion through out this paper it is sufficient to consider spatially one dimensional space. $\psi(q;t)$ is a complex-valued wave function supposed to determine the state of the particle. Let us project the above dynamics into real space. To do this, let us make the following transformation $\psi(q;t)=R(q;t)\exp(iS(q;t)/\hbar)$, where $R(q;t)$ and $S(q;t)$ are real-valued functions. Putting this into Eq. (\ref{Schroedinger equation}) and separating into the real and imaginary parts, one obtains
 \begin{eqnarray}
m\frac{dv}{dt}=-\partial_qU,\hspace{2mm}\partial_t\rho+\partial_q(v\rho)=0.
\label{Madelung fluid}
\end{eqnarray}
Here, $\rho(q)=|\psi(q)|^2=R^2(q)$ is quantum probability density, $U(q)$ is determined by the quantum amplitude $R(q)$ as 
\begin{equation}
U(q)= -\frac{\hbar^2}{2m}\frac{\partial_q^2R}{R},
\label{quantum potential}
\end{equation} 
and $v(q)$ is a velocity vector defined by the quantum phase $S(q)$ as
\begin{equation}
v(q)=\partial_qS/m.
\label{velocity field}
\end{equation} 

Equation (\ref{Madelung fluid}) to (\ref{velocity field}) is the so-called Madelung fluid picture of the Schr\"odinger equation \cite{Madelung paper}. Due to the formal similarity with Euler equation, the term on the right hand side of the left equation in Eq. (\ref{Madelung fluid}), $F\equiv-\partial_qU$,  is called as quantum force, thus accordingly, $U(q)$ is called as quantum potential. Putting in Madelung fluid form, it thus became explicit that the original Schr\"odinger equation possesses a hidden self-referential property. Namely, $U(q)$ is determined by $\rho(q)$ and in turn $U(q)$ will dictate the way $\rho(q)$ evolves with time through Eq. (\ref{Madelung fluid}) and so on and so forth. One thus may expect to observe self-organized physically interesting phenomena. 

Next, let us consider a class of quantum probability densities which maximizes Shannon information entropy: $H[\rho]=-\int dq\hspace{1mm}\rho(q)\ln\rho(q)$ \cite{Shannon entropy},  given its average quantum potential $\bar{U}=\int dq\hspace{1mm}U(q)\rho(q)$. This is the so-called maximum entropy principle \cite{Jaynes-MEP}. It has been argued as the only way to infer from an incomplete information which does not lead to logical inconsistency \cite{Shore-Johnson-MEP}. In our present case, the limited information that we have in hand is the average quantum potential. We shall show later that the average quantum potential is a physically relevant information. Hence, the maximum entropy principle will give us the most probable quantum probability density with average quantum potential $\bar{U}$. This inference  problem can be directly solved to give \cite{Mackey-MEP}:
\begin{equation}
\rho(q)=\frac{1}{Z}\exp\big(-U(q)/T\big),
\label{canonical QPD}
\end{equation}
where $T$ is a constant determined by ${\bar U}$, and $Z(T)$ is a normalization factor. Combining with Eq. (\ref{quantum potential}), Eq. (\ref{canonical QPD}) comprises a differential equation for $\rho(q)$ or $U(q)$, subjected to the condition that $\rho(q)$ must be normalized. In term of $U(q)$, one has \cite{AgungPRA1}
\begin{equation}
\partial_q^2U=\frac{1}{2T}(\partial_qU)^2+\frac{4mT}{\hbar^2}U. 
\label{NPDE for U}
\end{equation}

\begin{figure}[htbp]
\begin{center}
\includegraphics*[width=6cm]{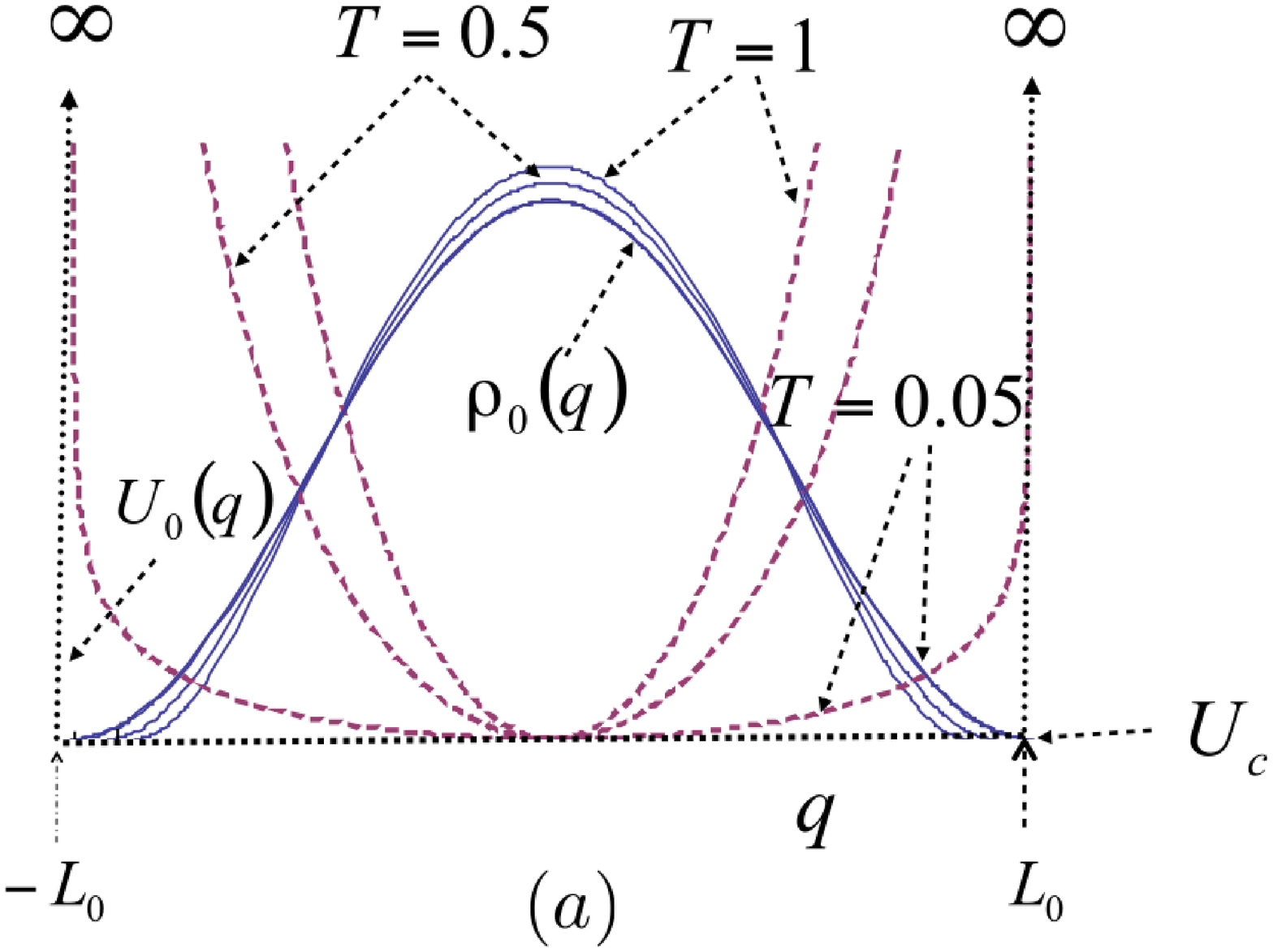}
\includegraphics*[width=7.5cm]{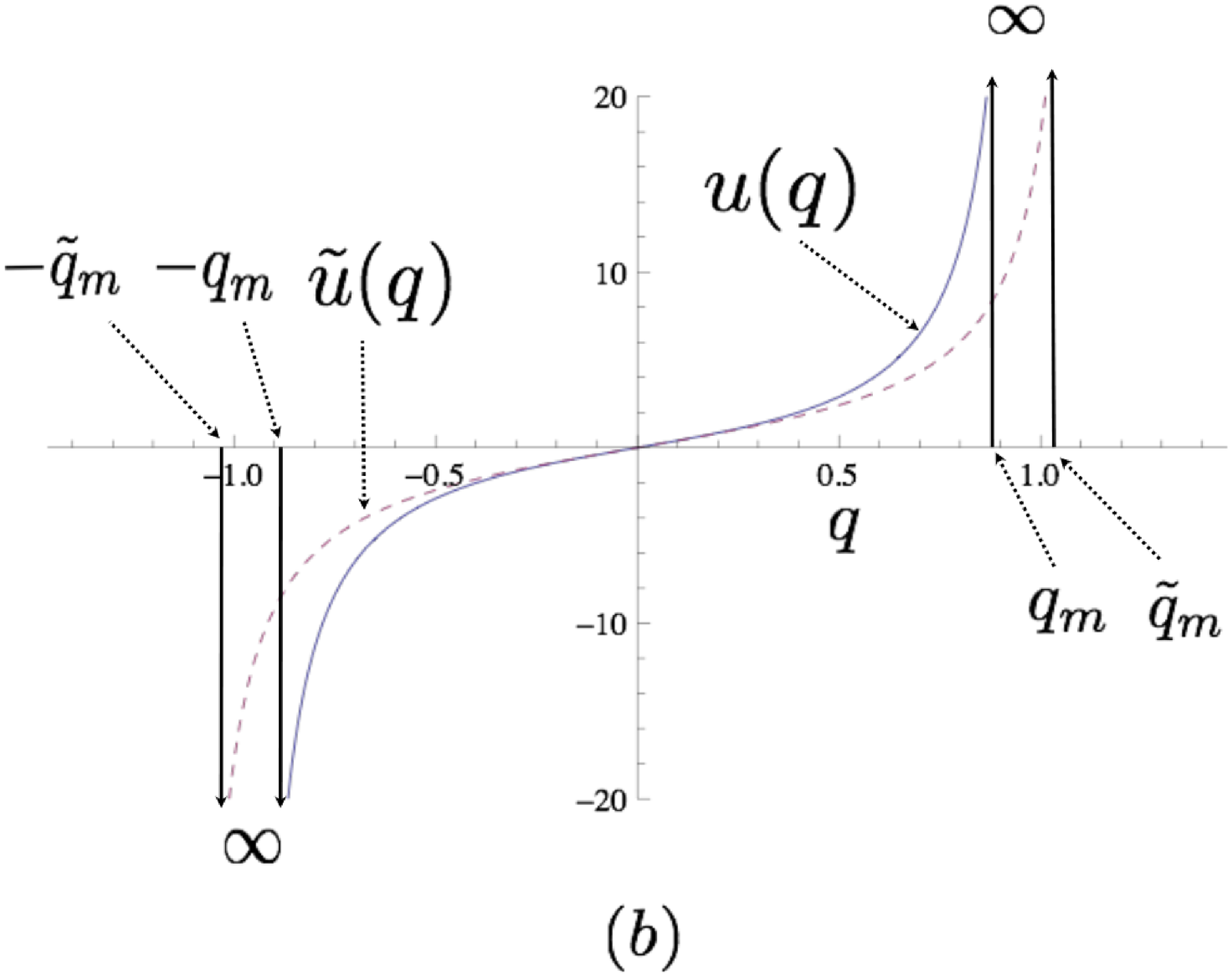}
\end{center}
\caption{(a) The quantum probability density (solid line) and its corresponding quantum potential (dashed line) for several small values of $T$ obtained by solving equation (\ref{NPDE for U}). We also plot the analytical solution for $\rho_0(q)$ at $T=0$, assuming that $U_0(q)$ takes the form of a box with infinite wall. See text for detail. (b) $u(q)$ and $\tilde{u}(q)$. See text for detail.}
\label{small temperature vs QPD and QP}
\end{figure}

Figure \ref{small temperature vs QPD and QP}a shows the solution of Eq. (\ref{NPDE for U}) with the boundary conditions: $U(0)=1$, $\partial_qU(0)=0$, for several small values of positive $T$. The reason for choosing small values of $T$ will be clear as we proceed. All numerical results in this paper are obtained by setting $\hbar=m=1$. One can see that the quantum probability density is being trapped by its own self-generated quantum potential. Moreover,  one can also see that there are points $q=\pm L_m$, where the quantum potential is blowing-up, namely $U(\pm L_m)=\infty$. This is a familiar phenomena in nonlinear differential equation \cite{blowing-up NDE} and in particular, for the case we are considering, it can be proven as follows. Let us define a new variable $u(q)=\partial_qU$. The nonlinear differential equation of Eq. (\ref{NPDE for U}) then transforms into 
\begin{equation}
\partial_qu=\frac{1}{2T}u^2+\frac{4mT}{\hbar^2}U.
\label{blowing-up NPDE for u}
\end{equation}
The boundary condition translates into $u(0)=\partial_qU(0)=0$. Further, let us now consider the following nonlinear differential equation
\begin{equation}
\partial_q\tilde{u}=\frac{1}{2T}\tilde{u}^2+\frac{4mT}{\hbar^2}X, 
\label{blowing-up for inferior solution}
\end{equation}
where $X\equiv U(0)$; with $\tilde{u}(0)=0$. Since $U(q)\ge U(0)=X$, then it is obvious that $|u(q)|\ge|\tilde{u}(q)|$. 

On the other hand, one can solve the latter nonlinear differential equation of Eq. (\ref{blowing-up for inferior solution}) analytically to have
\begin{equation}
\tilde{u}(q)=a\tan(bq), \hspace{2mm}a=\frac{2T}{\hbar}\sqrt{2mX}, \hspace{2mm}b=\frac{1}{\hbar}\sqrt{2mX}.  
\label{inferior solution}
\end{equation}
It is then clear that at $q=\pm\tilde{q}=\pm\pi/(2b)$, $\tilde{u}$ is blowing-up, namely $\tilde{u}(\pm\tilde{q})=\pm\infty$. Recalling the fact that $|u(q)|\ge|\tilde{u}(q)|$, then $u(q)$ is also blowing-up at points $q=\pm L_m$, $u(\pm L_m)=\pm\infty$, where $L_m\le \tilde{q}$. See Fig. \ref{small temperature vs QPD and QP}b. Hence, one can conclude that $U(q)$ is also blowing-up at $q=\pm L_m$, $U(\pm L_m)=\infty$. It is then safe to say that the self-trapped quantum probability density possesses only a finite range of spatial support $\mathcal{M}=[-L_m,L_m]$. 

\begin{figure}[htbp]
\begin{center}
\includegraphics*[width=7cm]{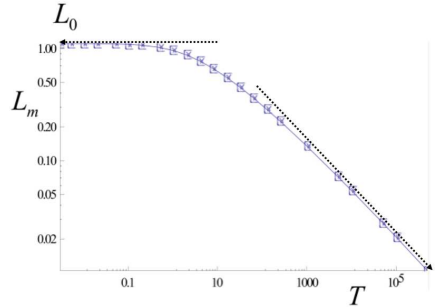}
\end{center}
\caption{The half length of the support of quantum probability density plotted against the parameter $T$.}
\label{rmax vs T}
\end{figure}

In Fig. \ref{rmax vs T}, we plot the variation of the half length of the support against  $T$. One observes that as $T$ increases toward infinity, $L_m(T)$ decreases toward zero. This shows that the corresponding quantum probability density collapses toward a delta function. On the contrary, as one decreases  $T$ toward zero, we see from Fig. \ref{rmax vs T} that $L_m$ is approaching to a finite value, $\lim_{T\rightarrow 0}L_m=L_0$. This hints us that the quantum probability density and thus its corresponding quantum potential are also converging toward some functions
\begin{eqnarray}
\lim_{T\rightarrow 0}U(q;T)=U_0(q),\hspace{2mm}\lim_{T\rightarrow 0}\rho(q;T)=\rho_0(q).
\label{QPD and QP at vanishing temperature}
\end{eqnarray}

To discuss this latter situation in more detail, let us go back to Fig. \ref{small temperature vs QPD and QP}a. One can see that as $T$ decreases, the quantum potential inside the support is getting flatterer before becoming infinite at the boundary points, $q=\pm L_m(T)$. One can thus guess that at vanishing $T$, $T=0$, the quantum potential is converging toward a box of length $2L_0$ with perfectly flat bottom and an infinite wall at the boundary points, $q=\pm L_0$. Let us show that our guess is correct. To do this, let us proceed to calculate the profile of the quantum probability density for vanishing $T$, $\rho_0(q)$, assuming that $U_0(q)$ takes a box potential of length $2L_0$. Namely $U_0(q)=U_c$, where $U_c$ is constant for $-L_0<q<L_0$, and $U_0(\pm L_0)=\infty$. Since $U_0(q)$ is flat inside the support, then one has $\bar{U}_0=\int_{-L_0}^{L_0} dq\rho_0(q)U_0(q)=U_c$. Hence, from the definition of quantum potential given in Eq. (\ref{quantum potential}), inside the support of the quantum probability density, one has 
\begin{equation}
-\frac{\hbar^2}{2m}\partial_q^2R_0=U_cR_0=\bar{U}_0R_0,
\label{stationary Schroedinger equation 1}
\end{equation} 
where $R_0(q)\equiv\rho_0^{1/2}(q)$ is the quantum amplitude at $T=0$. The above differential equation must be subjected to the boundary condition: $R_0(\pm L_0)=0$. 

Solving Eq. (\ref{stationary Schroedinger equation 1}) one obtains
\begin{equation}
R_0(q)=A_0\cos(k_0 q),
\label{localized-stationary QPD}
\end{equation}
where $A_0$ is a normalization constant and $k_0$ is related to the average quantum potential as 
\begin{equation}
k_0=\sqrt{2m\bar{U}_0/\hbar^2}. 
\label{wave number vs average quantum potential}
\end{equation}
The boundary condition imposes $k_0L_0=\pi/2$. In Fig. \ref{small temperature vs QPD and QP}a, we plot the above obtained quantum probability density, $\rho_0(q)$. One can see that as $T$ is decreasing toward zero, $\rho(q;T)$ obtained by solving the differential equation of Eq. (\ref{NPDE for U}) is indeed converging toward $\rho_0(q)$ given in Eq. (\ref{localized-stationary QPD}). This confirms our initial guess that at $T=0$ the quantum potential takes a form of box with infinite wall at $q=\pm L_0$. 

Let us remark that all the above results suggest that the wave function can generate its own boundary at which it is vanishing. This is a very important fact since then one can talk about a self-sustained geometrical property of the wave function. We shall show in the next section that this fact will help us in developing our notion of particle from the Schr\"odinger wave function. One can of course trace back the emergence of this self-organized structure from the self-referential property of the Madelung fluid dynamics. 

\section {Soliton wave packet} 

Let us now choose a pair of fields $\{\rho_0(q),v_0(q)\}$ as the initial state of the Madelung fluid. Here, $\rho_0(q)=R_0^2(q)$ is given in Eq. (\ref{localized-stationary QPD}) and $v_0(q)=v_c$ is a uniform velocity field with non-vanishing value only inside the spatial support $\mathcal{M}\equiv[-L_0,L_0]$. Then, since at $t=0$ the quantum potential is flat inside the support, the quantum force is initially vanishing, $\partial_qU=0$ such that $dv/dt=0$. Hence, the velocity field at infinitesimal lapse of time $t=\Delta t$ will stay unchanged and keeps uniform. This in turn will shift the quantum probability density in space by $\Delta q=v_c\Delta t$ while keeps its profile unchanged. Correspondingly, it will shift the support of the quantum probability density by the same amount. The same thing occurs for the next infinitesimal lapse of time and so on and so forth. One thus concludes that \textit{the quantum probability density is moving with a uniform velocity field $v_c$, keeping its initial form remained unchanged}. Hence, at time $t$, the quantum probability density is given by 
\begin{eqnarray}
\rho(q;t)=\rho_0(q-v_ct;0)=A_0^2\cos^2(k_0q-\omega_0 t),
\label{stationary-moving quantum probability density}
\end{eqnarray}
where $\omega_0=k_0v_c$ and $q\in \mathcal{M}_t\equiv[v_ct-L_0,v_ct+L_0]$. 

To write the explicit form of the complex-valued self-trapped and uniformly-moving yet stationary wave function $\psi(q;t)$ of our single particle, let us first calculate its quantum mechanical energy. Since quantum mechanical energy is conserved then it is sufficient to use the wave function at $t=0$. Putting the wave function in polar form, $\psi(q)=R_0(q)\exp(iS(q)/\hbar)$, one has 
\begin{eqnarray}
\langle E\rangle\equiv \int_{-L_0}^{L_0} dq\hspace{1mm}\psi^*(q)\Big(-\frac{\hbar^2}{2m}\partial_q^2\Big)\psi(q)\hspace{10mm}\nonumber\\
=\int_{-L_0}^{L_0} dq\hspace{1mm}\Big(-\frac{\hbar^2}{2m}R_0\partial_q^2R_0+\frac{1}{2m}R_0^2(\partial_qS)^2
\nonumber\\-\frac{i\hbar}{m}R_0\partial_qR_0\partial_qS-\frac{i\hbar}{2m}R_0^2\partial_q^2S \Big).
\label{calculation of quantum mechanical energy}
\end{eqnarray} 
The first term on the right hand side is equal to the average quantum potential, $\bar{U}_0=\int dqU_0\rho_0=U_c=\hbar^2k_0^2/(2m)$. Next, defining kinetic energy density as $K_0(q)\equiv(m/2)v_0^2(q)$, then the second term is equal to the kinetic energy of the Madelung fluid $\bar{K}_0=\int dq K_0(q)\rho_0(q)=(m/2)v_c^2$. Further, for a uniform velocity field, the last term is vanishing, $(1/m)\partial_q^2S=\partial_qv_0=0$. For the same reason, since $R_0(q)$ is an even function and $\partial_qR_0(q)$ is an odd function then the third term is also vanishing. 

Hence, in total, the quantum mechanical energy of the  self-trapped wave function moving with a uniform velocity field $v_c$ can be decomposed as 
\begin{equation}
\langle E\rangle=\bar{U}_0+\bar{K}_0=\frac{\hbar^2k_0^2}{2m}+\frac{mv_c^2}{2}.
\label{energy decomposition}
\end{equation}
Moreover, using similar argument as above, the average quantum momentum can also be calculated to give
\begin{eqnarray}
\langle p\rangle\equiv\int_{-L_0}^{L_0}dq\hspace{1mm}\psi^*(q)\big(-i\hbar\partial_q\big)\psi(q)=mv_c. 
\label{average momentum}
\end{eqnarray}
The average kinetic energy of the Madelung fluid and average quantum momentum are thus related as $\bar{K}_0=\langle p\rangle^2/(2m)$. This observation leads us to conclude that $\bar{U}_0$ must be interpreted as essentially an {\it internal energy} of the single particle. Namely it  is the energy when the particle is not moving. A similar notion is also developed in special relativity theory, dubbed as rest mass energy. One can also check that the internal energy is the energy which is missed to be taken into account if one uses the simple plane wave to represent a free moving particle. 

Bearing in mind the above derived facts, let us write the moving-stationary pair of fields $\{\rho(q),v(q)\}$ in complex-valued form, $\psi(q)=R(q;t)\exp(iS(q;t)/\hbar)$. To do this, one has to calculate the quantum phase $S(q;t)$ by integrating $v(q)=(1/m)\partial_qS=v_c$ to give $S(q;t)=mv_cq+\xi(t)$, where $\xi(t)$ depends only on time $t$. The wave function we are searching thus takes the following form: $\psi_s(q;t)=A_0\cos\big(k_0(q-v_ct)\big)\exp\Big(i\big(mv_cq+\xi(t)\big)/\hbar\Big)$. Finally, putting this back into the Schr\"odinger equation of Eq. (\ref{Schroedinger equation}), keeping in mind  $\omega_0=k_0v_0$ and Eq. (\ref{energy decomposition}), one easily obtains  $\partial_t\xi=-\langle E\rangle$, which can be integrated to give $\xi(t)=-\langle E\rangle\hspace{1mm} t$ modulo to some constant. We are thus led to the following form of wave function:
\begin{equation}
\psi_s(q;t)=A_0\cos\Big(k_0\Big(q-\frac{\langle p\rangle}{m}t\Big)\Big)\exp\Big(\frac{i}{\hbar}\big(\langle p\rangle q-\langle E\rangle t\big)\Big),
\label{stationary-moving wave function final}
\end{equation}
where $q\in\mathcal{M}_t=[\langle p\rangle t/m-L_0,\langle p\rangle t/m+L_0]$. The solitonic nature of the above wave function is obvious. 

\section {Compton wave length, phase resonance, and superposition} 

Let us proceed to discuss the stationary-moving soliton solution of the free particle Sch\"odinger equation that we just developed in the previous section. First, the wave function possesses only a finite range of support $\Delta q=2L_0$. Using the fact that $k_0L_0=\pi/2$ which came from the boundary condition for the quantum amplitude $R_0(L_0)=0$, and the expression for $k_0$ in term of quantum potential $U_c={\bar U}_0$ of Eq. (\ref{wave number vs average quantum potential}), one is led to the following relation between the width of the support of the soliton wave function and the quantum potential:
\begin{equation}
\Delta q=\sqrt{\frac{\hbar^2\pi^2}{2m{\bar U}_0}}.
\label{width vs internal energy}
\end{equation}

On the other hand, we have shown in the previous section that ${\bar U}_0$ is but the internal energy, namely the energy of the particle when it is not moving. It is thus reasonable to assume that this energy is equivalent to the rest mass energy predicted by the Einstein's theory of special relativity. Namely, one assumes that 
\begin{equation}
{\bar U}_0=mc^2,
\label{internal energy vs rest mass energy}
\end{equation}
where $c$ is the velocity of light. Putting this into Eq. (\ref{width vs internal energy}), one finally obtains
\begin{equation}
\Delta q=\frac{\pi}{\sqrt {2}}\frac{\hbar}{mc}=\frac{\pi}{\sqrt 2}\lambda_C,
\end{equation}
where $\lambda_C\equiv \hbar/(mc)$ is but the Compton wave length. One can thus conclude that the active  region of the soliton wave  function is proportional to the Compton wave length of the particle. In particular, this active region is narrower for particle with larger mass. 

\begin{figure}[htbp]
\begin{center}
\includegraphics*[width=9cm]{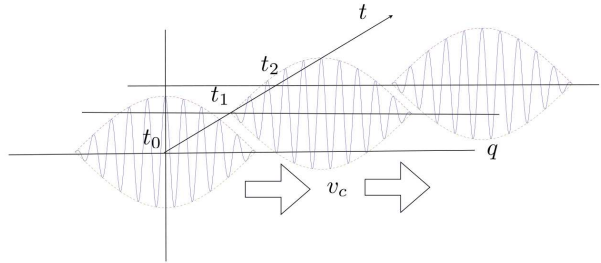}
\end{center}
\caption{The snapshot of the real part of soliton wave function, $\psi_s(q;t)$, at $t_0<t_1<t_2$.}
\label{linear soliton}
\end{figure}

Next, in Fig. \ref{linear soliton} we plot the snapshot of spatial profile of the real part of soliton wave function given in Eq. (\ref{stationary-moving wave function final}), $\psi_s(q;t)$, at times: $t_0<t_1<t_2$. One observes an envelope given by $R_0(q-v_ct)$ which is moving uniformly with velocity $v_c$. One can also see from Eq. (\ref{stationary-moving wave function final}), that there is {\it internal spatial vibration} whose wave number $k_B$ is determined by the average quantum momentum
\begin{equation}
k_B=\frac{\langle p\rangle}{\hbar}=\frac{mv_c}{\hbar}. 
\label{de Broglie wave length}
\end{equation}
$\lambda_B=2\pi/k_B=2\pi\hbar/(mv_c)$ is then but the famous de Broglie wave length. This is in fact what is expected by de Broglie in his conjecture of double solution. Yet in contrast to the theory of double solution which envisions a soliton solution emerging from a {\it nonlinear} wave equation, here we have shown a soliton wave function as a solution of {\it linear} Schr\"odinger equation. 

Before proceeding to discuss further consequence of the linearity of the Schr\"odinger equation, let us evaluate what will happen if the internal vibration of de Broglie's wave length resonates inside the self-generated quantum potential. Namely one imposes that the length of the support of the quantum probability density, $\Delta q=2L_0$, is equal to the integer multiple of the de Broglie's wave length $\lambda_B$. Hence, $k_B$ is equal to the integer multiple of $k_0$,  $k_B=nk_0$, where $n=0,\pm 1,\pm 2,\dots$. In this case, the velocity of the particle and the wave number of the envelope quantum amplitude $k_0$ are related as  
\begin{equation}
\langle p\rangle=mv_c=n\hbar k_0, \hspace{2mm}n=\pm 1,\pm 2,\dots. 
\label{principle of phase resonance}
\end{equation}
Namely, in this case the momentum of the particle can only take discrete possible values with the spacing given by $p_0=\hbar k_0$.  In the next section, we shall discuss a possible physical situation which requires the above phase resonance thus induces quantization.  

Finally, let us discuss important consequence of the linearity of the Schr\"odinger equation. Equation (\ref{stationary-moving wave function final}) tells us that given the energy $\langle E\rangle$ defined as in Eq. (\ref{energy decomposition}) and the momentum $\langle p\rangle$ of the single particle of mass $m$, one can associate a single soliton occupying a localized space moving with velocity $v=\langle p\rangle/m$. {\it Since the Schr\"odinger equation is linear, then any superposition of many soliton solutions of the type given in Eq. (\ref{stationary-moving wave function final}) will also satisfy the Schr\"odinger equation}. For example, let us consider the following superposition of two solitons solution
\begin{eqnarray}
\psi_{ms}(q;t)=A_1\cos\Big(k_1\big(q-v_1t\big)\Big)\exp\Big(\frac{i}{\hbar}\big(mv_1 q-E_1 t\big)\Big)\nonumber\\
+A_2\cos\Big(k_2\big(q-v_2t\big)\Big)\exp\Big(\frac{i}{\hbar}\big(mv_2 q-E_2 t\big)\Big). 
\label{two solitons solutions}
\end{eqnarray}
Each soliton term on the right hand side moves with uniform velocity $v_i$, $i=1,2$, and possesses a finite support given by $\mathcal{M}_{i}=[v_it-L_i,v_it+L_i]$, where $L_i=\pi/(2k_i)$. The total  support is then given by $\mathcal{M}_{ms}=\mathcal{M}_1\oplus\mathcal{M}_2$.  One can then show that the above wave function satisfies the Schr\"odinger equation of Eq. (\ref{Schroedinger equation}) if $E_i$ is given as follows:
\begin{equation}
E_i=\frac{\hbar^2k_i^2}{2m}+\frac{1}{2}mv_i^2,\hspace{2mm} i=1,2. 
\label{partial energy}
\end{equation}

Hence, it is as if each term on the right hand side of Eq. (\ref{two solitons solutions}) describes a de Broglie's particle of mass $m$ with internal energy ${\bar U}_i=\hbar^2k_i^2/(2m)$, moving with velocity $v_i$ thus possessing kinetic energy ${\bar K}_i=(1/2)mv_1^2$. Both of the particles are not interacting. Without loosing generality, let us consider the case when the sign of $v_1$ and $v_2$ are opposite to each other. First, at $t=0$, one observes spatial interference pattern. Hence, there is no way to distinguish one from the other. One thus must consider both as a single particle of mass $m$ with energy $\langle E\rangle$. Yet, at sufficiently large time, the two solitons are spatially separated, thus no interference phenomena is seen.  One can distinguish two spatially localized solitons corresponding to two particles. At this situation, it is easy to calculate the conserved quantum energy $\langle E\rangle$ to have
\begin{equation}
\langle E\rangle = E_1+E_2. 
\label{total energy of the two solitons}
\end{equation}
Namely, it is given as the summation of the energy of each particle, $E_i$. The theory then predicts an interesting phenomena in which a particle of a given mass $m$ can break into two particles with still equal mass $m$ yet possessing different energy. Since the Schr\"odinger equation is time reversal, then by reversing the time, one concludes that two particles with equal mass $m$ can merge into a particle of the same mass yet with energy given by the addition of the energy of the merging particles. 
 
One can of course extend the above two solitons solution to many solitons solution straightforwardly, each of which can be interpreted as describing a single particle of equal mass. Moreover, each moves independently from the other thus the total energy is given by the summation of the energy of all the particles.  Hence, the Sch\"odinger equation of Eq. (\ref{Schroedinger equation}) which previously is thought of as a theory of a single free particle, now is re-interpreted as a theory of non-interacting many particles with the same masses. 

\section{Conclusion and Discussion} 

First, we have shown that the linear Schr\"odinger equation for a single free particle possesses a class of soliton solution parameterized by the quantum energy and momentum. We showed that the soliton wave packet shares the properties that a quantum particle should have as envisioned by de Broglie long time ago in the early days of quantum mechanics even before the birth of the Schr\"odinger equation. Namely, the soliton wave function possesses an internal vibration whose wave number $k_B$ is determined by the momentum of the particle $\langle p\rangle$ exactly described by the famous de Broglie's relation, $\langle p\rangle=\hbar k_B$. 

We then showed that assuming resonance of the internal vibration inside the quantum potential, the possible values of the momentum of the particle is discretized as the integer multiple of the wave number of the envelope soliton. A natural question then arises: when this quantization happens? To answer this question, one can borrow conventional idea from the physics of classical wave. In this field, a resonance usually is related to interaction and transfer of energy. We therefore expect that the quantization in soliton wave function happens if it  interacts with other physical object. We shall elaborate this idea as future work. 

Finally, we showed that the linearity of the Schr\"odinger equation allows us to construct many solitons solution through the superposition principle. We argue that each term of the superposition can be  interpreted as describing a single particle with the same mass as the other, yet each might possesses different momentum and energy. Moreover, each particle is free to move independently from the other. Hence, we conclude that the Schr\"odinger equation of Eq. (\ref{Schroedinger equation}) which is conventionally thought of as a theory describing a single free particle, can now be regarded as a theory of many particles with equal mass. 

It is then interesting to further ask: can we have a similar theory which possesses many solitons solution so that each can be associated to a particle with different masses? To do  this one can not start from a theory in which the mass is already fixed as in the Schr\"odinger equation of Eq. (\ref{Schroedinger equation}). Namely, one has to develop a theory in which the masses emerge as consequence of the theory. This theory should be based on a mass-less wave equation.  We shall report this some where else.

\end{document}